\documentclass{jps-cp}
\usepackage{txfonts} 
\usepackage{graphicx}
\usepackage{xcolor}

\usepackage{amsmath}
\usepackage{xurl}

\title{ Time-dependent weighted directed networks of cryptocurrency interaction from high-frequency returns}

\author{Shubhangam \textsc{Shukla}$^{1}$, Mahesh \textsc{Peyyala}$^{2}$ and Abhijit \textsc{Chakraborty}$^{1,3,*}$}

\inst{$^{1}$Department of Humanities and Social Sciences, Indian Institute of Science Education and Research Tirupati, Srinivasapuram, Yerpedu Mandal, Tirupati, Andhra Pradesh, 517619, India  \\
$^{2}$Department of Physics, Indian Institute of Science Education and Research Tirupati, Srinivasapuram, Yerpedu Mandal, Tirupati, Andhra Pradesh, 517619, India  \\
$^{3}$RIKEN Center for Interdisciplinary Theoretical and Mathematical Sciences, Saitama, 351-0198, Japan}

\email{* abhijit@labs.iisertirupati.ac.in\\
}

\recdate{April 21, 2026}

\abst{

We investigate the evolving structure of interactions in cryptocurrency markets using a network-based framework constructed from high-frequency price data spanning 2020-2025. Directed and weighted networks are constructed from statistically significant Granger causal relationships between cryptocurrency log-returns, enabling us to quantify the flow of influence across assets. We find that normalized returns exhibit heavy-tailed distributions, consistent with the presence of large intermittent fluctuations and in line with stylized facts of financial markets. The resulting networks display pronounced heterogeneity in link weights and nodal strengths, indicating that a small subset of cryptocurrencies contributes disproportionately to market dynamics. By ranking cryptocurrencies based on their nodal out-strength, we uncover a dynamically evolving hierarchy of influence. Ethereum consistently emerges as the most influential asset, while Bitcoin shows a gradual decline in its relative importance. The ranking structure exhibits substantial temporal variability, with multiple cryptocurrencies entering and exiting the top positions over time. Our findings reveal a highly competitive and non-stable organization of the cryptocurrency ecosystem.
}
\kword{cryptocurrencies, log-return, granger causality, weighted networks}

\begin{document}
\maketitle

\section{Introduction}
Cryptocurrencies have emerged as a prominent class of digital financial assets, characterized by decentralized governance, high market liquidity, and continuous trading. Unlike traditional financial markets, cryptocurrency markets operate around the clock and exhibit pronounced price fluctuations across multiple time scales. These fluctuations are often quantified using logarithmic returns, which reveal heavy-tailed distributions and intermittent extreme events, indicative of complex underlying dynamics~\cite{easwaran2015bitcoin, beguvsic2018scaling, kakinaka2020characterizing}. 

Beyond individual price behavior, cryptocurrencies are not isolated entities but form an interacting system where information and shocks propagate across assets. Such interactions arise from shared market sentiment, arbitrage opportunities, and common external influences, leading to correlated and potentially causal relationships among different cryptocurrencies. These interdependencies can be effectively captured using network-based approaches, where nodes represent individual assets and links encode statistical or causal interactions, such as those inferred from Granger causality analysis~\cite{scagliarini2022pairwise}.
In addition to price-based networks, recent studies have also examined networks constructed from wallet-level transaction data, where nodes represent users or addresses and links correspond to fund transfers, revealing structural properties such as clustering, centralization, and the emergence of influential entities in the ecosystem~\cite{chakraborty2023projecting, chakraborty2023dynamic, chakraborty2023embedding, chakraborty2024arbitrage, ikeda2023hodge}. The resulting networks provide complementary insights into both market dynamics and user-level activity, identifying dominant cryptocurrencies and influential participants, as well as revealing the collective organization and temporal evolution of the system. The study of cryptocurrency price fluctuations and their interactions offers a valuable framework for understanding the complex behavior of modern financial systems, as also observed for  traditional financial markets~\cite{laloux1999noise, plerou1999universal} and the foreign-exchange market~\cite{chakraborty2018deviations, chakraborty2020uncovering}.

In this work, we adopt a network-based framework to investigate the temporal evolution of interactions in the cryptocurrency market using high-frequency data from 2020 to 2025. We construct time-dependent directed and weighted networks in which nodes represent cryptocurrencies and links are inferred from statistically significant Granger causal relationships between their price fluctuations~\cite{scagliarini2022pairwise, granger1969, hamilton2020time}. Our analysis shows that normalized log-returns exhibit broad, heavy-tailed distributions, indicating the presence of large intermittent fluctuations and reinforcing the universality of stylized facts across financial systems. 

The inferred networks display strong heterogeneity in both link weights and nodal strengths, with a small subset of cryptocurrencies contributing disproportionately to the overall flow of influence. By ranking nodes according to their out-strength, we uncover a dynamically evolving hierarchy of influence, where leading cryptocurrencies such as Ethereum maintain sustained dominance, while others, including Bitcoin, exhibit gradual changes in their relative importance. Notably, the ranking structure shows significant temporal variability, with multiple assets entering and exiting the top positions over time, in contrast to the persistence of super-stable nodes reported in other complex systems~\cite{ghoshal2011ranking}. These findings highlight the competitive and non-stationary nature of the cryptocurrency ecosystem and demonstrate the effectiveness of complex network approaches in uncovering its underlying organization.

The remainder of this paper is structured as follows. Section~\ref{data} introduces the dataset and its preprocessing. Section~\ref{method} outlines the methodological framework employed in this study. The main findings are presented and discussed in detail in Section~\ref{result}. Finally, Section~\ref{conclusion} summarizes the results and provides concluding remarks.

\section{Data}
\label{data}

We construct time-dependent networks of cryptocurrencies, where nodes represent individual assets and directed links are inferred from causal relationships between their price fluctuations. The dataset is obtained from the Kraken Crypto Exchange~\cite{Data}, which provides publicly available trade-level information, including timestamps (with second-level resolution), traded prices in USD and other fiat currencies, and transaction volumes. Throughout this study, we use USD as the reference currency due to its relative stability.

To obtain uniformly sampled time series, prices are aggregated at a one-minute resolution using the volume-weighted average price (VWAP), while the traded volume within each minute is computed as the sum of all transactions. For time intervals with no recorded trades, the last observed price is carried forward to ensure continuity, and the corresponding volume is set to zero. Price fluctuations are quantified using logarithmic returns defined as
\begin{equation}
R(t) = \log P(t+\Delta t) - \log P(t),
\end{equation}
where $P(t)$ denotes the price at time $t$ and $\Delta t = 1$ minute.

The resulting time series are divided into nonoverlapping weekly windows, each consisting of $10{,}080$ data points (corresponding to seven days at one-minute resolution). The analysis spans the period from January 2020 to March 2025, providing a total of $275$ weekly windows. Over this period, the number of actively traded cryptocurrencies increases substantially, from approximately $30$ at the beginning of 2020 to about $390$ by March 2025.

Prior to executing the Granger causality tests, we systematically verified the stationarity of the logarithmic return time series for every asset across all weekly windows using the Augmented Dickey--Fuller (ADF) test~\cite{dickey1979distribution} at a significance level of $\alpha = 0.01$. Under the null hypothesis that a time series contains a unit root (i.e. is non-stationary), our analysis confirms that $99.06\%$ of the evaluated data are strictly stationary. The remaining $0.94\%$ represents periods of extreme trading inactivity characterized by stale pricing and absolute zero variance (i.e., mathematically flat logarithmic returns). Because Vector Autoregression (VAR) estimation strictly requires dynamic price variation, these inactive instances inherently lack predictive power and were systematically excluded from pairwise evaluations for their respective weeks.

\section{Methods}
\label{method}

To infer interactions between cryptocurrencies, we employ the Granger causality (GC) framework, which quantifies the extent to which past values of one time series improve the prediction of another~\cite{granger1969, scagliarini2022pairwise}. For each pair of stationary time series $(X,Y)$ of log-returns, we consider two linear autoregressive models: a restricted model and a full model.

In the restricted model, the target variable $Y$ is expressed solely in terms of its own past values:
\begin{equation}
Y_t = \sum_{m=1}^{p} a_m\, Y_{t-m} + \varepsilon'_t,
\label{eq:restricted}
\end{equation}
where $\varepsilon'_t$ denotes the residual error. 

In the full model, past values of the source variable $X$ are additionally included:
\begin{equation}
Y_t = \sum_{m=1}^{p} a_m\, Y_{t-m}
      + \sum_{m=1}^{p} b_m\, X_{t-m} + \varepsilon_t,
\label{eq:full}
\end{equation}
where $\varepsilon_t$ is the corresponding residual. The optimal lag order $p$ is determined by minimizing the Bayesian information criterion (BIC)~\cite{schwarz1978estimating} using maximum lag = 10.

The strength of Granger causality from $X$ to $Y$ is quantified as
\begin{equation}
G_{X \to Y} = \ln\!\left(\frac{\sigma_r^2}{\sigma_f^2}\right),
\label{eq:gcstrength}
\end{equation}
where $\sigma_r^2 = \langle \varepsilon_t^{\prime 2} \rangle$ and $\sigma_f^2 = \langle \varepsilon_t^{2} \rangle$ denote the variances of the residuals from the restricted and full models, respectively. We note that Granger causality is a statistical concept, quantifying the extent to which past values of one time series improve the linear prediction of another. It does not necessarily imply a direct physical or mechanistic causal relationship between the assets.

A directed link from $X$ to $Y$ is established if the estimated GC strength exceeds a significance threshold.  The resulting networks are therefore directed and weighted, with link weights corresponding to the magnitude of the Granger causality strength~\cite{scagliarini2022pairwise, hamilton2020time}. This procedure is applied independently within each weekly window, yielding a sequence of evolving networks over the study period. 

Since Granger causality is tested independently for every ordered pair of cryptocurrencies within each weekly window, the number of simultaneous hypothesis tests grows quadratically with the number of assets, substantially increasing the risk of false positives under a fixed significance threshold. To control for this multiple-testing problem, we apply the Benjamini--Hochberg (BH) false discovery rate (FDR) procedure~\cite{benjamini1995controlling} to the $p$-values associated with all pairwise GC tests within each weekly window. Specifically, the $p$-values $\{p_{(1)}, p_{(2)}, \dots, p_{(n)}\}$ obtained from the $n = N(N-1)$ pairwise tests (for $N$ cryptocurrencies) are first sorted in ascending order. A link from $X$ to $Y$ is then declared significant if its $p$-value satisfies
\begin{equation}
p_{(k)} \leq \frac{k}{n}\, q,
\label{eq:bhfdr}
\end{equation}
where $k$ is the rank of the $p$-value in the sorted sequence and $q$ is the desired FDR level, set to $q = 0.05$ in this study. All links satisfying Eq.~\eqref{eq:bhfdr} are retained as statistically significant, while the remaining links are discarded.  The weight of the retained link is then set to the log-variance ratio $G_{X \to Y}$ defined in Eq.~\eqref{eq:gcstrength}.  This correction is applied independently within each weekly window prior to network construction, ensuring that the expected proportion of false positives among the retained directed links remains controlled at the specified FDR level across the entire study period.
 
\begin{figure}[t]
\centering
\includegraphics[width=0.98\textwidth]{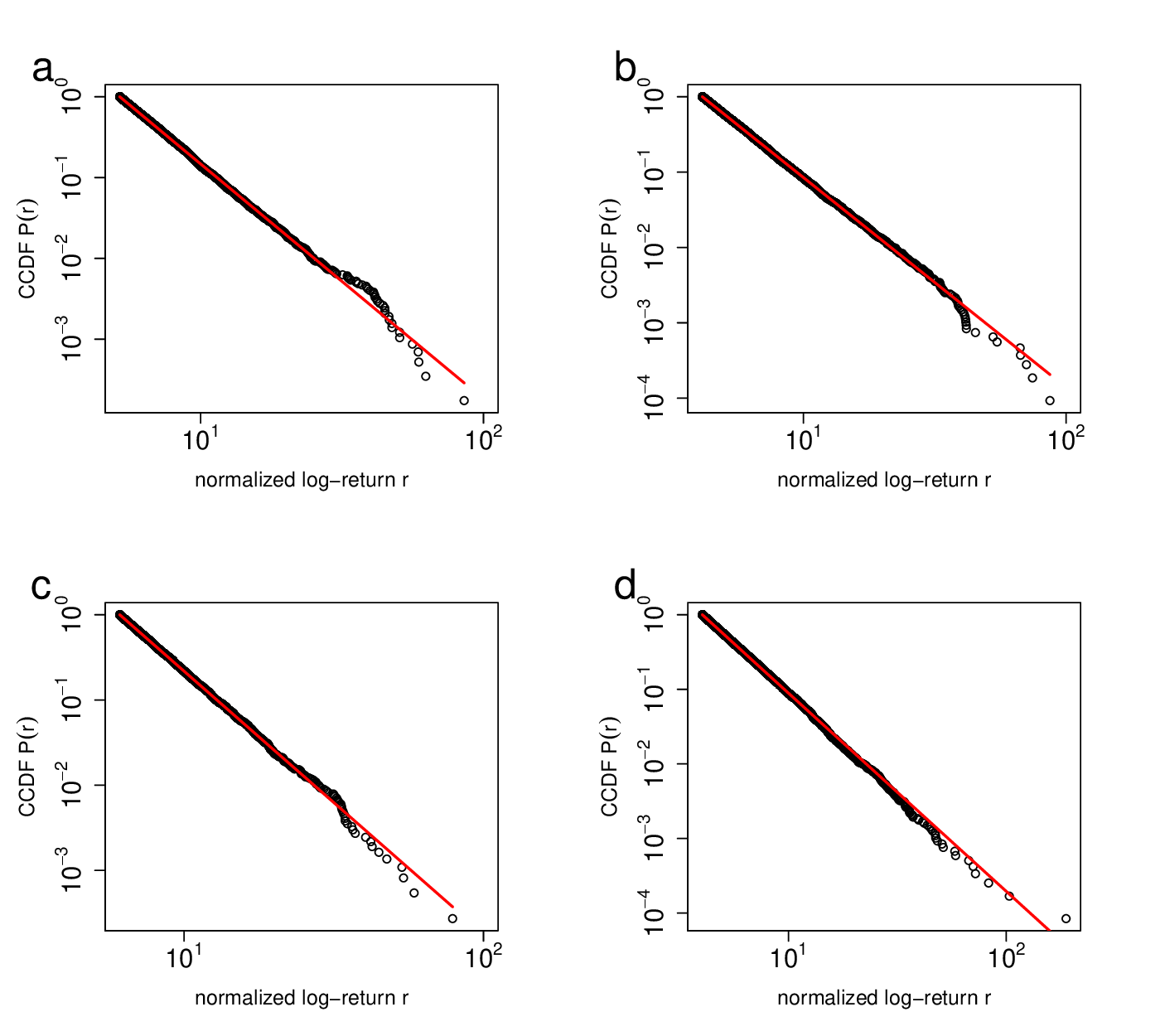}
\caption{Complementary cumulative distribution functions (CCDF) of the normalized 1-minute log-returns $r$. Panels (a) and (b) show the positive and negative tails for XBT, respectively, while panels (c) and (d) correspond to XRP. The red lines represent power-law fits of the form $P(r) \sim r^{1-\gamma}$, with estimated exponents $\gamma = 3.91, 3.78, 4.08,$ and $3.66$ for panels (a)--(d), respectively. The exponents are obtained using the maximum likelihood estimation method.}
\label{f1}
\end{figure}
\section{Results}
\label{result}
We show the complementary cumulative distribution functions (CCDF) of the normalized 1-minute log-returns for selected cryptocurrencies over the period $2020$--$2025$ in Fig.~\ref{f1}. The normalized log-return, $r_i$, which quantifies price fluctuations for $i$-th cryptocurrency, is defined as
$
r_i(t) = (R_i(t) - \langle R_i \rangle)/\sigma_i,
$
where $\langle R_i \rangle$ and $\sigma_i$ denote the mean and standard deviation of $R_i$, respectively, computed over the full sample.
 As evident from the figure, the fluctuations exhibit a broad range, extending approximately within $[-100, 100]$ for XBT and XRP, indicating significant variability in short-time price dynamics. Other cryptocurrencies display qualitatively similar behavior.

A detailed analysis of the tails of the CCDF distributions reveals that both the positive and negative tails follow a power-law form, $P(r) \sim r^{1-\gamma}$. Using maximum likelihood estimation, we obtain the tail exponents $\gamma = 3.91$ and $3.78$ for the positive and negative tails of XBT, and $\gamma = 4.08$ and $3.66$ for XRP, respectively. These values indicate heavy-tailed behavior, characteristic of large, intermittent fluctuations in financial time series.

Interestingly, an earlier study on XBT fluctuations~\cite{easwaran2015bitcoin} reported a smaller tail exponent of approximately $3$. The observed deviation in our results may be attributed to structural changes in the cryptocurrency market, which has evolved and matured significantly since its early stages. Notably, similar power-law behavior has been widely reported in traditional financial markets, including stock prices~\cite{gopikrishnan1998inverse, pan2008inverse} and foreign exchange rates~\cite{chakraborty2018deviations}, suggesting a degree of universality in the statistical properties of price fluctuations across different asset classes.
\begin{figure}[t]
\centering
\includegraphics[width=0.99\textwidth]{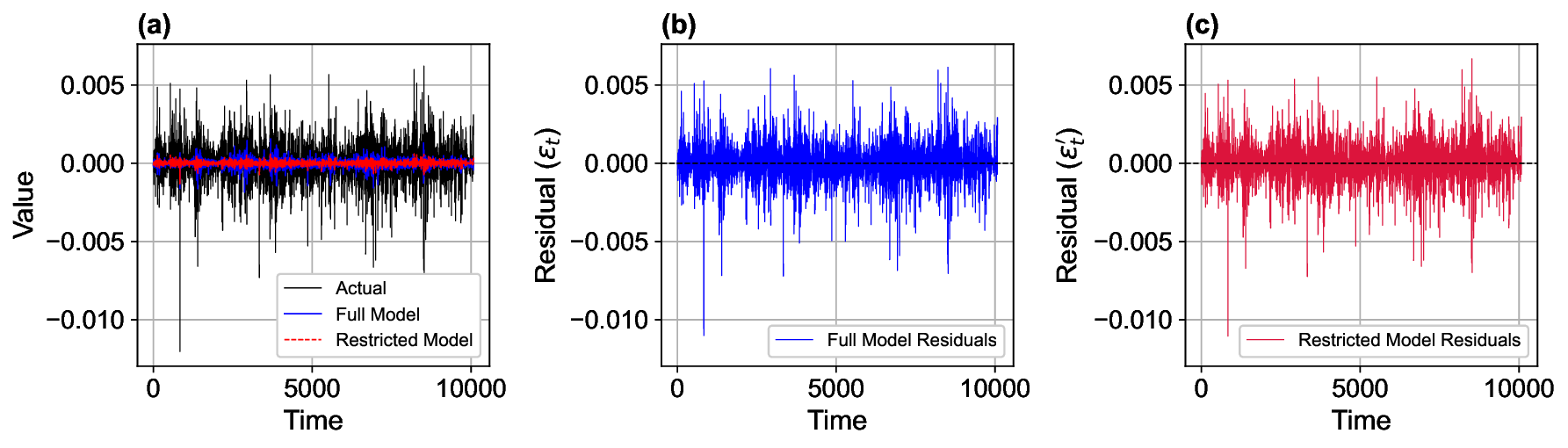}   
\caption{(a) 1-minute log-return of ETH (grey), along with the restricted model (red) defined in Eq.~\ref{eq:restricted}, where the target variable $Y$ represents log-return time series for ETH, and the full model (blue) defined in Eq.~\ref{eq:full}, where $X$ represents log-return time series for XBT. In both models, the error term is excluded. Panels (b) and (c) show the corresponding residuals (error terms) for the full model and restricted model, respectively.}
\label{f2}
\end{figure}

We illustrate the estimation of Granger causality between a pair of cryptocurrencies, following the methodology described in Sec.~\ref{method} in Fig.~\ref{f2}. As an example, we compute the Granger causality from XBT to ETH, $G_{{\rm XBT} \to {\rm ETH}}$, using their 1-minute log-return time series $R (t)$ for the week of January $6$--$12$, 2020. In Fig.~\ref{f2}(a), the empirical log-return of ETH is fitted using both the restricted model [Eq.~\ref{eq:restricted}] and the full model [Eq.~\ref{eq:full}]. The corresponding residuals (error terms) are shown in Fig.~\ref{f2}(b) for the full model and in Fig.~\ref{f2}(c) for the restricted model.

A summary of the estimation of $G_{{\rm XBT} \to {\rm ETH}}$ is provided in Table~\ref{t1}. The results indicate that including XBT in the full model for predicting ETH reduces the variance of the residuals by $3.24\%$ compared to the restricted model. This reduction in prediction error suggests that XBT Granger-causes ETH during the considered time period.
\begin{table}[tbh]
\caption{Summary of the Granger causality estimation from XBT to ETH, $G_{{\rm XBT} \to {\rm ETH}}$.}
\label{t1}
\centering
\begin{tabular}{lc}
\hline\hline
\textbf{Metric} & \textbf{Value} \\
\hline
Source & XBT \\
Target & ETH \\
Optimal lag & $2$ \\
Full model variance & $7.2 \times 10^{-7}$ \\
Restricted model variance & $7.5 \times 10^{-7}$ \\
Variance reduction & $3.24\%$ \\
$G_{{\rm XBT} \to {\rm ETH}}$ & $0.033$ \\
F-statistic & $171.4$ \\
$p$-value & $< 10^{-10}$ \\
\hline\hline
\end{tabular}
\end{table}

\begin{figure}[t]
\centering
\includegraphics[width=0.98\textwidth]{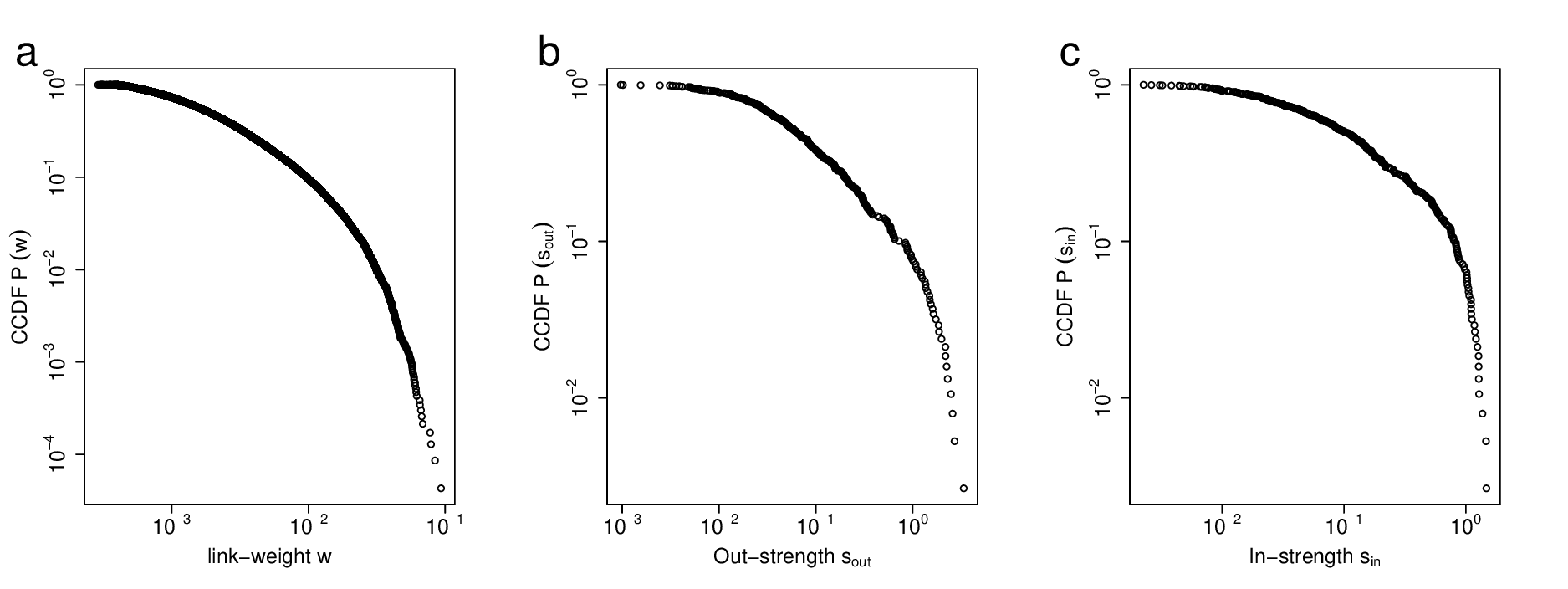}
\caption{Complementary cumulative distribution functions (CCDF) of key network measures: (a) link weight $w$, (b) nodal out-strength $s_{\mathrm{out}}$, and (c) nodal in-strength $s_{\mathrm{in}}$.
Results are shown for a representative week (24--30 March 2025); other weeks exhibit qualitatively similar behaviour.
}
\label{f3}
\end{figure}

We construct a weighted, directed network of cryptocurrencies by estimating pairwise Granger causality based on 1-minute logarithmic return series $R(t)$. From these interactions, we generate a sequence of weekly networks spanning the period from 2020 to 2025. We show the complementary cumulative distribution functions (CCDF) of key weighted network quantities in Fig.~\ref{f3}. Panel (a) presents the distribution of link weights $w$, while panels (b) and (c) show the distributions of nodal out-strength $s_{\mathrm{out}}$ and in-strength $s_{\mathrm{in}}$, respectively. All three distributions exhibit broad, heavy-tailed behavior, indicating significant heterogeneity in both the strength of interactions and the distribution of influence across nodes. In particular, the heavy tails in $s_{\mathrm{out}}$ and $s_{\mathrm{in}}$ suggest that a small number of nodes contribute disproportionately to the total outgoing and incoming influence in the network, reflecting an uneven and highly skewed structure of interactions.

\begin{figure}[t]
\centering
\includegraphics[width=0.65\textwidth]{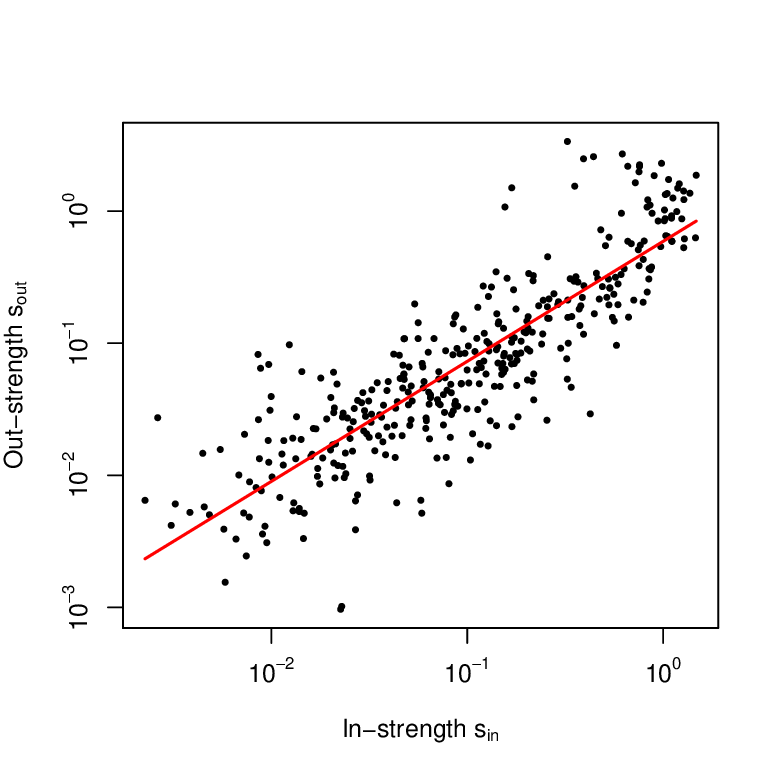}
\caption{Scatter plot of nodal out-strength $s_{\mathrm{out}}$ versus in-strength $s_{\mathrm{in}}$ for all nodes. The red line represents a power-law fit of the form $s_{\mathrm{out}} \sim s_{\mathrm{in}}^{~\alpha}$. The estimated exponent $\alpha = 0.91$ indicates a nontrivial sublinear relationship between the two quantities. Results are shown for a representative week (24--30 March 2025); other weeks exhibit qualitatively similar behaviour.}
\label{f4}
\end{figure}

Figure~\ref{f4} shows the relationship between the nodal out-strength $s_{\mathrm{out}}$ and in-strength $s_{\mathrm{in}}$ for all nodes in the network. Each point represents a node, and the data are displayed on logarithmic scales to highlight scaling behavior over several orders of magnitude. A clear positive correlation between $s_{\mathrm{out}}$ and $s_{\mathrm{in}}$ is observed, indicating that nodes with higher incoming strength also tend to have higher outgoing strength.

To quantify this relationship, we fit the data with a power-law of the form $s_{\mathrm{out}} \sim s_{\mathrm{in}}^{~\alpha}$. The estimated exponent $\alpha = 0.909$ reveals a nontrivial sublinear scaling between the two quantities. This sublinear behavior suggests that although highly connected nodes are influential both in terms of incoming and outgoing interactions, the increase in outgoing strength is slower than linear with respect to the incoming strength, reflecting an inherent asymmetry in the flow of influence across the network. A log-log linear regression was performed to examine the relationship between in-strength ($s_{in}$) and out-strength ($s_{out}$) of the network nodes. The fitted model, $\log_{10}(s_{out}) = \beta_0 + \beta_1 \log_{10}(s_{in})$, yielded a highly significant positive relationship between the two variables ($\beta_1 = 0.909$, SE $= 0.029$, $t(376) = 31.65$, $p < 2 \times 10^{-16}$), with an intercept of $\beta_0 = -0.229$ (SE $= 0.035$, $t(376) = -6.62$, $p = 1.25 \times 10^{-10}$). The model explained a substantial proportion of the variance in out-strength ($R^2 = 0.727$, adjusted $R^2 = 0.726$), and the overall regression was highly significant ($F(1,376) = 1002$, $p < 2.2 \times 10^{-16}$). To determine whether the scaling exponent deviates from linearity, a one-sample $t$-test was conducted against the null hypothesis $\beta_1 = 1$, yielding $t(376) = (0.909 - 1)/0.029 = -3.18$, $p = 0.0016$. The corresponding 95\% confidence interval for $\beta_1$, $[0.852,\ 0.965]$, excludes unity, confirming that the scaling exponent is statistically significantly less than 1. This indicates a genuinely sublinear relationship between in-strength and out-strength, whereby nodes with higher incoming connectivity exhibit disproportionately lower outgoing connectivity strength, rather than the two scaling proportionally.

To check whether liquidity could explain the observed scaling---highly traded assets may naturally both transmit and receive more information, simultaneously inflating their in- and out-strength---we re-ran the regression with the total weekly traded volume $V$ of each asset included as an additional covariate. The extended model,
\begin{equation}
\log_{10}(s_{out}) = \beta_0 + \beta_1\,\log_{10}(s_{in}) + \beta_2\,\log_{10}(V),
\label{eq:liquiditycontrol}
\end{equation}
yields $\beta_1 = 0.903$ (SE $= 0.029$, $t(374) = 31.24$, $p < 2\times10^{-16}$), virtually identical to the estimate from the bivariate model. The volume coefficient is statistically insignificant ($\beta_2 = 0.016$, SE $= 0.012$, $t(374) = 1.35$, $p = 0.178$), indicating that traded volume provides no additional explanatory power once in-strength is accounted for. The model fit is unchanged ($R^2 = 0.726$, adjusted $R^2 = 0.725$; $F(2,374) = 496.5$, $p < 2.2\times10^{-16}$). These results confirm that the sublinear scaling between in-strength and out-strength is a genuine structural property of the inferred networks and is not confounded by asset liquidity.

\begin{figure}[t]
\centering
\includegraphics[width=0.99\textwidth]{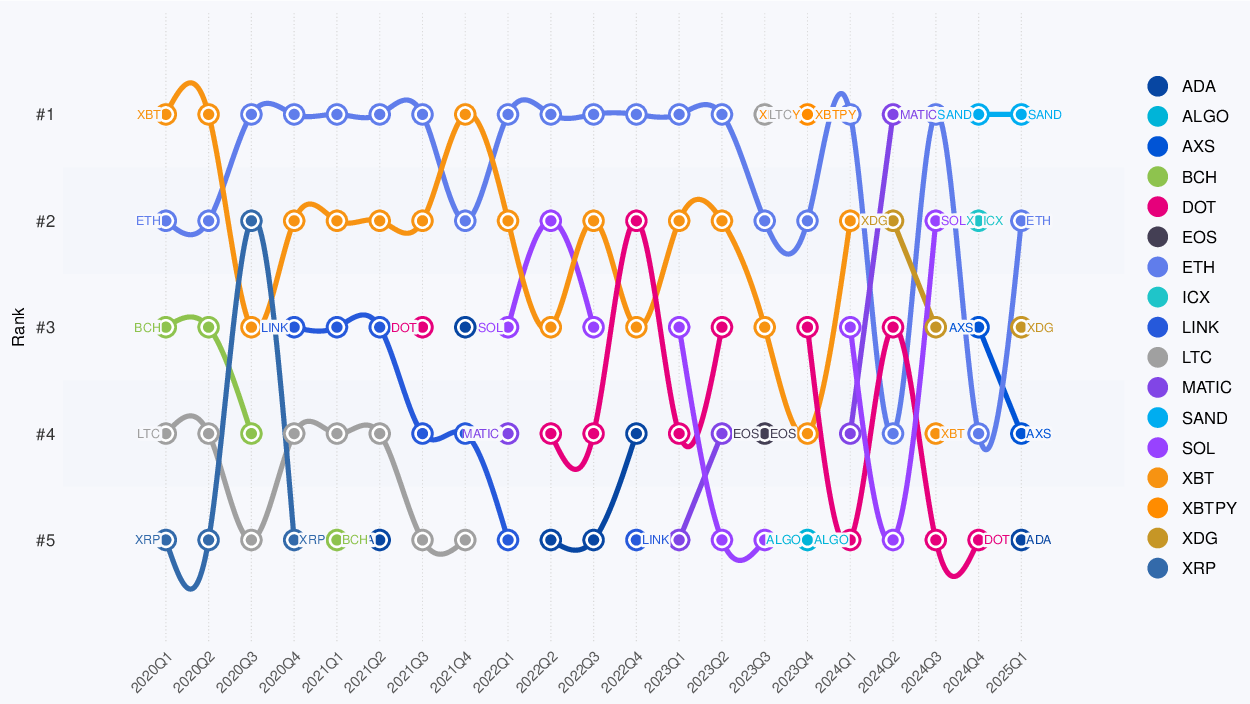}
\caption{Ranking of cryptocurrencies based on their average quarterly out-strength (influence) over the period 2020--2025. Ethereum (ETH) consistently ranks as the most influential cryptocurrency, while Bitcoin (XBT) shows a gradual decline in influence. The rankings exhibit significant temporal variability, particularly between Q3 2022 and Q1 2025, with multiple cryptocurrencies entering and exiting the top ranks.
The symbols are defined as follows: \texttt{XBT} (Bitcoin), \texttt{XBTPY} (Bitcoin Paxos, a tokenized form of Bitcoin settled on the Paxos blockchain), \texttt{ETH} (Ethereum), \texttt{XRP} (Ripple), \texttt{XDG} (Dogecoin), \texttt{ADA} (Cardano), \texttt{SOL} (Solana), \texttt{DOT} (Polkadot), \texttt{LTC} (Litecoin), \texttt{LINK} (Chainlink), \texttt{BCH} (Bitcoin Cash), \texttt{MATIC} (Polygon), \texttt{ALGO} (Algorand), \texttt{EOS} (EOS), \texttt{AXS} (Axie Infinity), \texttt{SAND} (The Sandbox), and \texttt{ICX} (ICON).
}
\label{f5}
\end{figure}
We present the temporal ranking of cryptocurrencies based on their average quarterly out-strength in Fig.~\ref{f5}, which serves as a measure of their overall influence in the network. To calculate the average quarterly out-strength, we first normalize the out-strength of each node by the total number of nodes, $N$, in the network. The quarterly measure is then obtained by averaging the normalized out-strength values over all weekly networks belonging to the same quarter.
Ethereum (ETH) consistently occupies the top position throughout the period 2020--2025, indicating its sustained and stable influence.  ETH's persistent dominance in the rankings may reflect ongoing technological developments within the Ethereum ecosystem, including its landmark transition to a proof-of-stake consensus mechanism, which strengthened confidence in the network's future growth and sustainability.
In contrast, Bitcoin (XBT) exhibits a gradual decline in its relative influence over time. 
Ripple (XRP) holds the \#3 position in the early period but is gradually displaced by Cardano (ADA), Solana (SOL), and other emerging assets from 2021 onward. This decline may be partly attributable to increased regulatory uncertainty surrounding Ripple. In December 2020, the U.S. Securities and Exchange Commission (SEC) initiated legal proceedings against Ripple Labs and two of its executives, alleging violations of U.S. securities laws. The lawsuit introduced significant uncertainty regarding the regulatory status of XRP and may have contributed to its subsequent loss of market prominence. Solana shows a particularly sharp ascent, climbing into the Top~5 by 2023--2024 and maintaining its position thereafter. In contrast, Bitcoin Cash (BCH), EOS, Litecoin (LTC), and Algorand (ALGO) make brief appearances in the Top~5 during the early quarters but largely disappear from the rankings after 2021, reflecting a broader shift in market capitalization toward newer, high-growth platforms. The ranking dynamics reveal pronounced temporal variability, especially during the period from Q3 2022 to Q1 2025, where the composition of top-ranked cryptocurrencies changes frequently. One possible explanation for the growing instability in cryptocurrency rankings is the collapse of FTX in November 2022. The failure of the exchange, previously valued at over $32$ billion, exposed significant governance and risk-management shortcomings, leading to widespread market uncertainty and a reassessment of asset valuations across the cryptocurrency ecosystem.
In total, 17 distinct cryptocurrencies appear at least once among the top five most influential assets over the five-year span. This observation highlights the absence of persistently dominant nodes and stands in contrast to earlier findings on real-world networks that report the existence of super-stable nodes~\cite{ghoshal2011ranking}. The results thus suggest a highly dynamic and competitive structure of influence in the cryptocurrency market.

\begin{figure}[t]
\centering
\includegraphics[width=0.99\textwidth]{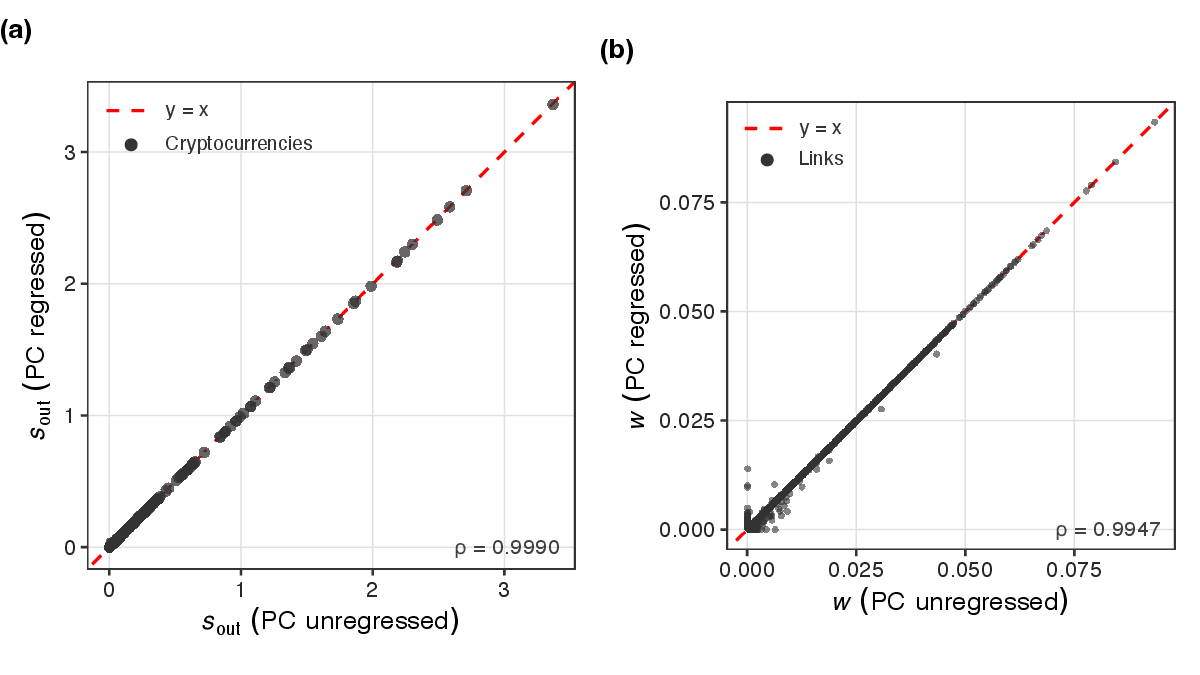}
\caption{Robustness of the inferred network structure after removing the common
    market factor via first principal component (PC1) regression. Panel~(a) compares
    the nodal out-strength $s_{out}$ computed from the original log-returns (PC
    unregressed) against those computed from the PC1-filtered idiosyncratic residuals
    (PC regressed) for all nodes across the network. Panel~(b) shows the corresponding
    comparison for individual Granger causality link weights $w$. In both panels, the
    red dashed line indicates the identity $y = x$. The Spearman rank correlations
    between the original and PC1-filtered quantities are $\rho = 0.9990$ for the
    out-strengths and $\rho = 0.9947$ for the link weights, confirming that the
    network structure is preserved after controlling for shared market-wide
    co-movement. Results are shown for a representative week (24--30 March 2025); other weeks exhibit qualitatively similar behaviour.
}
\label{f6}
\end{figure}

To verify that the inferred Granger causal links reflect genuine pairwise asset-to-asset influence rather than spurious co-movement driven by a shared market-wide shock, we repeated the full network construction after removing the first principal component (PC1) of the return matrix from each asset's time series. For each weekly window, we construct the log-return matrix $R \in \mathbb{R}^{T \times N}$ and compute the $N \times N$ covariance matrix of the assets. The leading eigenvector $\mathbf{w}_1$, corresponding to the largest eigenvalue, captures the direction of maximum co-variance across the market, and the projection
\begin{equation}
M = R\mathbf{w}_1 \in \mathbb{R}^{T \times 1}
\label{eq:marketmode}
\end{equation}
defines the market mode time series. Each asset's log-return $R_{i,t}$ is then regressed against $M_t$ via ordinary least squares,
\begin{equation}
R_{i,t} = \alpha_i + \beta_i M_t + \varepsilon_{i,t},
\label{eq:olsresidual}
\end{equation}
where $\alpha_i$ is the intercept, $\beta_i$ measures the asset's sensitivity to the global market mode, and $\varepsilon_{i,t} = R_{i,t} - (\alpha_i + \beta_i M_t)$ is the idiosyncratic residual that is orthogonal to the common factor. The pairwise Granger causality analysis described in Sec.~\ref{method} is then applied to these residual series $\{\varepsilon_{i,t}\}$ in place of the raw returns.
 
Figure~\ref{f6}~(a) compares the nodal out-strengths $s_{out}$ obtained from the original and PC1-filtered networks across all nodes and all weekly windows. The data points cluster tightly around the $y = x$ line over the entire range, demonstrating that the distribution of influence among nodes is essentially unaffected by the removal of the common market component. Figure~\ref{f6}~(b) shows the corresponding comparison at the level of individual Granger causality link weights $w$. The bulk of links again fall on the $y = x$ diagonal, with only minor deviations in a small subset of weak links. The Spearman rank correlation between the two out-strength distributions is $\rho =0.9990$ and between the two link weight distributions is $\rho =0.9947$, confirming near-perfect preservation of the network structure. Taken together, these results confirm that the network structure, including both the nodal strength and the link weight distribution, is robust to the presence of shared market-wide shocks, and that the inferred Granger causal relationships capture genuine pairwise asset-to-asset interactions.

\section{Conclusions}
\label{conclusion}

In this work, we have analyzed the structure and evolution of interactions in cryptocurrency markets using a network-based approach grounded in Granger causality. By constructing time-dependent directed and weighted networks from high-frequency price data over the period 2020--2025, we have characterized the flow of influence among cryptocurrencies and its temporal variability.

Our results show that cryptocurrency price fluctuations exhibit heavy-tailed distributions, reflecting the presence of large and intermittent market movements. The constructed interaction networks display strong heterogeneity in both link weights and nodal strengths, indicating that influence is unevenly distributed across assets. In particular, a small subset of cryptocurrencies accounts for a disproportionately large share of the total outgoing and incoming influence, highlighting the presence of dominant yet evolving hubs in the system.

The analysis of nodal strength reveals a nontrivial sublinear relationship between out-strength and in-strength, suggesting an inherent asymmetry in how influence is distributed and transmitted across the network. Furthermore, by ranking cryptocurrencies based on their out-strength, we uncover a dynamically changing hierarchy of influence. Ethereum consistently maintains a leading position throughout the study period, while Bitcoin exhibits a gradual decline in its relative influence. The composition of top-ranked cryptocurrencies is highly time-dependent, with frequent turnover and the absence of persistently dominant nodes, in contrast to the super-stable structures reported in other complex systems.

The present study extends prior work on Granger-causality-based cryptocurrency networks, most notably that of Scagliarini~\textit{et al.}~\cite{scagliarini2022pairwise}, along several key dimensions. While Ref.~\cite{scagliarini2022pairwise} constructs weekly cryptocurrency trading networks using log-returns restricted to the years 2020--2021 and reports a network structure and set of influential nodes that remain comparatively stable over that short window, the present analysis employs log-returns spanning a substantially longer period (January 2020 to March 2025), yielding $275$ weekly networks and allowing the universe of traded assets to grow from approximately $30$ to nearly $390$ coins. This combination of higher temporal resolution and longer time span, together with careful pre-processing---we test every series for stationarity using the Augmented Dickey--Fuller test and correct for the multiple-testing problem inherent to pairwise Granger causality using the Benjamini--Hochberg false discovery rate procedure---allows us to reconstruct the directed influence networks with greater statistical rigor and longer temporal detail than earlier studies. Methodologically, whereas Ref.~\cite{scagliarini2022pairwise} primarily characterizes the trading network through pairwise Granger causality and high-order (O-information) dependencies to assess overall network stability, the present work focuses on the nodal out-strength as a long-horizon, quarterly-resolved measure of influence, uncovering a markedly different and more dynamic picture: rather than a stable hierarchy dominated by a persistent set of coins, we find pronounced turnover in the top-ranked assets, with 17 distinct cryptocurrencies occupying the top five positions at some point over the five-year span, and a clear transition of dominance from Bitcoin to Ethereum. This longer-horizon, higher-resolution perspective additionally reveals a previously unreported nontrivial sublinear scaling relationship between nodal in-strength and out-strength ($s_{out} \sim s_{in}^{\alpha}$, $\alpha = 0.909$), pointing to an inherent asymmetry in the propagation of influence that was not captured in earlier weekly-resolution, shorter-horizon studies of the cryptocurrency market.

Overall, our findings highlight the intrinsically dynamic and competitive nature of the cryptocurrency market, where influence is continuously redistributed among assets. The network-based framework employed here provides a powerful tool for uncovering the structure of interactions and identifying key drivers of market behavior. Future work may extend this approach by incorporating higher-order dependencies, nonlinear causality measures, or combining price-based networks with transaction-level (wallet) networks to obtain a more comprehensive understanding of the cryptocurrency ecosystem.



%


\section*{Acknowledgements} We thank Tomas Scagliarini for helpful comments on multiple hypothesis testing.

\end{document}